\documentclass[showpacs,aps,prd,twocolumn,nofootinbib]{revtex4}
\usepackage{amsmath}
\usepackage{epsfig}
\usepackage{subfigure}

\newcommand{\ep}{\varepsilon}

\newcommand{\eqs}[1]{\begin{equation} \begin{split} #1\end{split} \end{equation} }

\newcommand{\ga}{\gamma^5}
\newcommand{\gmu}{\gamma^\mu}

\newcommand{\gsig}{\gamma^\sigma}

\newcommand{\ie}{{\it i.e.}}

\newcommand{\etal}{{\it et al.}}

\newcommand{\ce}[1]{Eq.~(\ref{#1})}

\newcommand{\ct}[1]{{Table~(\ref{#1})}}

\begin{document}

\bigskip
\title{Two-photon width of $\eta_b$, $\eta_b'$ and $\eta_b''$ from Heavy-Quark Spin Symmetry}

\author
{J.P. Lansberg$^{a,b}$ and T.N. Pham$^{a}$}

\affiliation{
$^{a}$Centre de Physique Th\'eorique, \'Ecole polytechnique, CNRS,  \\
91128 Palaiseau, France
\\$^{b}$Physique Th\'eorique Fondamentale, Universit\'e de Li\`ege,\\
17 All\'ee du 6 Ao\^ut, B\^at. B5, B-4000 Li\`ege-1, Belgium}

 
\begin{abstract}
\parbox{14cm}{\rm 
We predict the two-photon width of the pseudoscalar bottomonia, 
$\Gamma_{\gamma \gamma}(\eta_b)$, $\Gamma_{\gamma \gamma}(\eta'_b)$ and
$\Gamma_{\gamma \gamma}(\eta''_b)$
within a Heavy-Quark Spin-Symmetry setting following the same line as
our recent work for the corresponding decays for
charmonia. Binding-energy effects 
are included for excited states and are shown to shift up the 
$\eta'_b$ and $\eta''_b$ widths by 
10 \%. We point out that the essentially model independent ratio 
$\Gamma_{\gamma \gamma}(\eta_b)/\Gamma_{\ell \bar \ell}(\Upsilon)$
and the branching ratio $BR(\eta_{b} \to \gamma\gamma)$ obtained 
could be used to extract the coupling constant $\alpha_{s}$.}
\end{abstract}
\maketitle

\section{Introduction}

In this brief report, we calculate the two-photon 
width of the pseudoscalar bottomonia, $\Gamma_{\gamma \gamma}(\eta_b(nS))$ 
using the same straightforward approach as in~\cite{Lansberg:2006dw} 
which employs  an effective Lagrangian satisfying 
heavy-quark spin symmetry~\cite{neubert,nardulli,DeFazio}. Since the 
$b$-quark mass is significantly
higher than the  $c$-quark mass, we expect the approximations
in  our approach to hold better.

For the time being, only a single candidate for 
the   ground state ($\eta_b$) has been found by the 
Aleph collaboration~\cite{Heister:2002if}, its mass
was evaluated to be $9300 \pm 28$ MeV~\cite{Heister:2002if,PDG}. 
The Potentiality for its observation at the Tevatron was also analysed
in~\cite{Maltoni:2004hv}.

On the contrary, all the corresponding  vector states below the $B\bar B$ threshold have
been observed and their leptonic-decay width measured. Indeed, the $\Upsilon(1S)$ 
mass is $9460.30 \pm 0.26$ MeV and its leptonic width $\Gamma_{\ell \bar\ell}$ is $1.340 \pm 0.018$ keV, 
$M_{\Upsilon(2S)}$ is $10.02326 \pm 0.00031$ GeV, $\Gamma_{\ell \bar \ell}(\Upsilon(2S))$ is 
$0.612 \pm 0.011$ keV, 
 $M_{\Upsilon(3S)}$ is $10.3552 \pm 0.0005$ GeV  and finally 
$\Gamma_{\ell \bar\ell}(\Upsilon(3S))$ is $0.443 \pm 0.008$ keV~\cite{PDG}. For recent reviews
on new mesons, see~\cite{Swanson:2006st,Rosner:2006jz,Colangelo:2006aa}.

From these widths, we can extract $f_{\Upsilon(nS)}$. With  
$f_{\eta_b(nS)}\approx f_{\Upsilon(nS)}$ --following 
Heavy-Quark Spin Symmetry (HQSS)--  we can predict
the corresponding two-photon decay rates of the $\eta_b(nS)$. For the 
$2S$ and $3S$ states, we 
are also able to refine the procedure by including binding-energy effects 
 as a function of the to-be measured $\eta_{b}'$ and  $\eta_{b}''$ masses. 
Moreover these decay constants 
could also be computed via sum rules~\cite{Reinders:1982zg} or lattice 
simulations~\cite{Dudek:2006ej}.

As in~\cite{Lansberg:2006dw}, the effective Lagrangians, which we use for the coupling 
of the $b \bar b $ pair to 
two photons and to a dilepton pair $\ell \bar \ell$, are:
\eqs{
{\cal L}^{\gamma \gamma}_{\rm eff}=&-i c_1(\bar b \gsig \ga b) \ep_{\mu \nu \rho \sigma} 
F^{\mu\nu} A^\rho,\\
{\cal L}^{\ell \bar \ell}_{\rm eff}=&-  c_2(\bar b \gmu b) (\ell \gamma_\mu \bar \ell),
}
with $\displaystyle c_1\simeq \frac{Q_b^2 (4\pi
  \alpha_{em})}{M_{\eta_b}^2+b_{\eta_b} M_{\eta_b}}$ 
 and 
$\displaystyle c_2=\frac{Q_b (4\pi \alpha_{em})}{M_\Upsilon^2}$. 

The factor $1/(M_{\eta_b}^2+b_{\eta_b} M_{\eta_b})$ in $c_{1}$ 
contains the binding-energy effects~\cite{Kuhn:1979bb,Pham} 
and is obtained from the denominator of the bottom-quark 
propagator. $b_{\eta_b}$ $(=2m_{b} -M_{\eta_b})$ is the bound-state binding energy
and $M_{\cal Q}$ its mass (in order to be consistent, 
we keep only terms linear in $b_{\eta_b}$, since the $O(q^{2}/m_{b}^{2})$ terms
have been neglected in the propagator). Since $\Lambda_{\rm QCD}\ll m_{b}$
the above effective Lagrangians should work better for  bottomonia than 
for  charmonia.

Defining $ \left<0|\bar b \gmu  b| \Upsilon \right>\equiv f_{\Upsilon}
M_\Upsilon\ep^\mu$, similarly to~\cite{Lansberg:2006dw}, we have for the
leptonic width 
\eqs{\label{eq:lept_width}
\Gamma_{\ell \bar \ell}(\Upsilon)=\frac{1}{64 \pi^2 M_\Upsilon} \int d \Omega |{\cal M}|^2=\frac{4 \pi Q_b^2 \alpha^2_{em} f_\Upsilon^2}{3 M_\Upsilon}.
}
Using $M_{\Upsilon}f_\Upsilon^2=12 |\psi(0)|^2$~\cite{Novikov:1977dq}, the non-relativistic result of
Kwong~\etal~\cite{Kwong:1987mj} is recovered. 

Similarly,  with $ \left<0|\bar b \gmu \ga b| \eta_b \right>\equiv i f_{\eta_b} P^\mu$, the  $\eta_b(nS)$ width into two photons is readily obtained:
\eqs{\label{eq:2ph_width}
\Gamma_{\gamma \gamma}(\eta_b)=\frac{1}{2}\frac{1}{64 \pi^2 M_{\eta_b}} \int d \Omega |{\cal M}|^2=\frac{4 \pi Q_b^4 \alpha^2_{em} M_{\eta_b}^{3} f_{\eta_b}^2}{(M_{\eta_b}^2+b_{\eta_b} M_{\eta_b})^{2}},
}
the factor $\frac{1}{2}$ being the Bose-symmetry factor.

NLO corrections are taken into account thanks to~\cite{Kwong:1987mj}
\eqs{\label{eq:NLO-QCD}
\Gamma^{NLO}(^3S_1)&= \Gamma^{LO} \left(1- \frac{\alpha_s}{\pi}\frac{16}{3}\right), \\
\Gamma^{NLO}(^1S_0)&= \Gamma^{LO} \left(1- \frac{\alpha_s}{\pi}\frac{(20-\pi^2)}{3}\right).}

Apart from predictions of decay rates, it is also very fruitful to consider ratios of
decay rates, since in some interesting cases radiative corrections
cancel out --up to a shift of the renormalisation scale due to different
bottomonium masses, that we can safely neglect. 

Indeed, a very convenient way~\cite{Lansberg:2006dw} to calculate decay width of 
excited-pseudoscalar states ($\eta'_b$ and $\eta''_b$) is to use the following relation
derived from \ce{eq:lept_width}, \ce{eq:2ph_width} and HQSS:
\eqs{\label{eq:g_etabp}
\Gamma_{\gamma \gamma}(\eta_b(nS))
=\Gamma_{\gamma \gamma}(\eta_b)
\left(\frac{M_{\eta_b}^2+b_{\eta_b} M_{\eta_b}}{M_{\eta_b(nS)}^2+b_{\eta_b(nS)} 
M_{\eta_b(nS)}}\right)^2 
 \\\times  \frac{M_{\eta_b(nS)}^3}{M_{\eta_b}^3}
\left(\frac{\Gamma_{e^+e^-}(\Upsilon(nS))}{\Gamma_{e^+e^-}(\Upsilon(1S))}
  \frac{M_{\Upsilon(nS)}}{M_{\Upsilon(1S)}}\right),}
where the bracket in the second line is  equivalent to $\frac{f^2_{\Upsilon(nS)}}{f^2_{\Upsilon(1S)}}$. Its main 
asset is to provide with predictions independent of radiative corrections as soon as
$\Gamma_{\gamma \gamma}(\eta_b)$, $\Gamma_{e^+e^-}(\Upsilon(nS))$ and 
$\Gamma_{e^+e^-}(\Upsilon(1S))$ are known.

We can also consider the ratio of $\Gamma_{\gamma \gamma}(\eta_b)$ over 
$\Gamma_{\ell \bar \ell}(\Upsilon)$. For the ground state $\eta_b$, 
the effect of the binding energy $b_{\eta_b}$, which depends on the pole mass $m_{b}$ (around $4.75$ GeV~\cite{Lakhina:2006vg}),
is small and can be neglected. Including NLO corrections, 
 from \ce{eq:NLO-QCD}, we find:

\eqs{\label{eq:R} R_{\eta_{b}} = \frac{\Gamma_{\gamma \gamma}(\eta_b)}{\Gamma_{\ell \bar \ell}(\Upsilon)}=
3\,Q_b^2\,\frac{M_\Upsilon}{M_{\eta_b}}\left(1+\frac{\alpha_s}{\pi}\frac{(\pi^{2}- 4)}{3}\right)}

Finally, we can compute the branching ratio into two photons which is given~\cite{Kwong:1987mj} by   
\eqs{\label{eq:BR}  \frac{\Gamma_{\gamma\gamma}(\eta_b)}{\Gamma_{gg}(\eta_{b})}=
\frac{9}{2}\,Q_b^4\,\frac{\alpha^{2}_{em}}{\alpha^{2}_{s}}\left(1- 7.8\,\frac{\alpha_s}{\pi}\right).}
This relation is expected to give a good prediction for the 
$\eta_{b}\to \gamma\gamma$ branching ratio since a corresponding  expression 
for $\eta_{c}$ gives the $\eta_{c}\to \gamma\gamma$ branching ratio 
in good agreement with experiment.

\section{Results}

Inserting the experimental value of 
$\Gamma_{e^+ e^-}(\Upsilon)$, $1.340 \pm 0.018$ keV, as well as its mass, in~\ce{eq:lept_width}, we get 
an estimate of $f_{\Upsilon}$, $836$ MeV, taking into account NLO corrections 
thanks to \ce{eq:NLO-QCD} (we have set $\alpha_s(M_{\Upsilon})=0.16$). 
For excited  $\Upsilon$ states, we correspondingly get $f_{\Upsilon'}$ at $583$ MeV and $504$ MeV for the  $\Upsilon''$.

As suggested by HQSS, 
we suppose the equality between $f_{\Upsilon}$ and
$f_{\eta_b}$; from~\ce{eq:2ph_width} with $b_{\eta_b(1S)}\simeq 0$, we
obtain the following  evaluation, $\Gamma_{\gamma \gamma}(\eta_b)=560$
eV ($\alpha_s(M_{\Upsilon})=0.16$, $M_{\eta_b}=9300$ MeV) . 

Extrapolating HQSS to $2S$ states,~\ie~$f_{\Upsilon(2S)}=f_{\eta'_b}$ and
$f_{\Upsilon(3S)}=f_{\eta''_b}$ , and neglecting
mass effects ($b=0$, $M_{\cal Q}=2m_b$), we obtain from~\ce{eq:g_etabp}
$\Gamma^{}_{\gamma \gamma}(\eta'_b)= \Gamma^{}_{\gamma \gamma}(\eta_b) 
\frac{f^2_{\Upsilon(2S)}}{f^2_{\Upsilon(1S)}}=250~\hbox{eV}$ as well as
$\Gamma^{}_{\gamma \gamma}(\eta''_b)= \Gamma^{}_{\gamma \gamma}(\eta_b) 
\frac{f^2_{\Upsilon(3S)}}{f^2_{\Upsilon(1S)}}=187~\hbox{eV}$.

To take into account binding-energy effects,
 we need to evaluate $b_{\eta_b(2S)}$ and
$b_{\eta_b(3S)}$ and thus a prediction of their mass. We shall use
$10.00$ GeV for the $\eta_b(2S)$ and $10.35$ GeV for the $\eta_b(3S)$ following  
the predictions of~\cite{Godfrey:1985xj,Lakhina:2006vg}.

Inserting our evaluation of $\Gamma^{}_{\gamma \gamma}(\eta_b)$ in~\ce{eq:g_etabp}, we eventually obtain
$\Gamma^{}_{\gamma \gamma}(\eta'_b)=269~\hbox{eV}~\&~\Gamma^{}_{\gamma \gamma}(\eta''_b)=
208~\hbox{eV}$.
A variation of $\pm 25$ MeV in the mass of $\eta_b(2S)$ and $\eta_b(3S)$ only affect our prediction
by 1-2 eV and it is therefore negligible. We conclude that the introduction of 
mass effects has shifted up the widths by about 20 eV. 

Again, in \ce{eq:g_etabp}, the radiative corrections  are cancelled up to corrections 
due to differences in the scale of $\alpha_s$ and can be also used to predict 
the width of $\eta_b(2S)$ and $\eta_b(3S)$ once an experimental value 
for $\Gamma_{\gamma \gamma}(\eta_b(1S))$ is available.

Another point underlying the importance of measuring the two-photon 
width of $\eta_{b}$ lies in the fact that the ratio of the  
two-photon decay rate of $\eta_{b}$ to the $\Upsilon$ leptonic width,
as given by the quantity $R_{\eta_b}$ in \ce{eq:R}, depends essentially on 
the QCD coupling constant $\alpha_{s}$. Since HQSS is found to work for 
$\eta_{c}$ and since it is expected to work better for $\eta_{b}$, 
this relation could be used to determine in a reliable way 
the value of $\alpha_{s}$ for the process. Given the
measured value of $\alpha_{s}$ at $M_{Z}$, the momentum scale at which  
$\alpha_{s}$ is to be evaluated here  could be in principle be fixed 
with $R_{\eta_b}$. This ratio was also considered in the context of  the
nonrelativistic renormalisation group approach at NNLL~\cite{Penin:2004ay}. 
For a recent review on pNRQCD, see~\cite{Brambilla:2004jw}.

As done in the past with the leptonic and hadronic branching ratio 
for $\Upsilon$~\cite{Mackenzie,Kwong:1987mj} ,
one can further check the consistency
of the value for $\alpha_{s}$ by considering also the branching
ratio for $\eta_{b}\to \gamma\gamma$ which is given by \ce{eq:BR}.

\begin{widetext}

\begin{table}[h]
\begin{tabular}{cccccccccccc}
\hline\hline 
$\Gamma_{\gamma \gamma}$ &  This paper &Sch.~\cite{Schuler:1997yw} &Lak.~\cite{Lakhina:2006vg}&Ack.~\cite{Ackleh:1991dy}& Kim~\cite{Kim:2004rz}&
Ahm.~\cite{Ahmady:1994qf} & M\"un.~\cite{Munz:1996hb} &Eb.~\cite{Ebert:2003mu}&God.~\cite{Godfrey:1985xj}&Fab.~\cite{Fabiano:2002nc}&Pen.~\cite{Penin:2004ay}\\ 
\hline
$\eta_b$ &   $560$ & $460$& $230$&$170$ & $384 \pm 47$& $520$ &$220 \pm 40$ & $350$ &$214$&
$466\pm 101$&$659\pm92$
\\
$\eta'_b$ &  $269$ & $200$& $70$ & - & $191 \pm 25$& - & $110 \pm 20$& 150& 121&-&-\\
$\eta''_b$&  $208$ & -    & $40$ & - & - & -& $84 \pm 12$& 100 & 90.6& -&-\\
\hline\hline
\end{tabular}
\caption{Summary of theoretical predictions for 
$\Gamma_{\gamma \gamma}(\eta_b)$, $\Gamma_{\gamma \gamma}(\eta'_b)$ and 
$\Gamma_{\gamma \gamma}(\eta''_b)$. (All values
are in units of eV).}\label{tab-res}
\end{table}

\end{widetext}

\section{Conclusion}

In~\cite{Lansberg:2006dw}, we have recently stressed that there may exist a specificity in the two-photon
decay of the first radially-excited pseudoscalar charmonium $\eta'_c$ whereas
it is usually the production of heavy-quarkonium  which creates the greatest debates
(for recent reviews, see~\cite{yr,Lansberg:2006dh}). More specifically, 
we have shown that the inclusion of the mass effects for $\eta'_c$ --supposedly
 important for excited states-- worsens the comparison between theoretical
expectation and the recent measurement by the CLEO collaboration~\cite{Asner:2003wv}.

In this work, we have evaluated the two-photon width of the pseudoscalar bottomonia
$\eta_b(nS)$ through the simple application of HQSS. 
For $n>1$ states, we
have taken into account binding-energy corrections, or equally speaking mass effects. 
They amount to up to 10 \% 
and our prediction depend little on the value of their mass. 
As HQSS and local operator expansion 
are  supposed to hold better for ground-state bottomonia,
the forthcoming measurement of $\Gamma_{\gamma \gamma}(\eta_b(1S))$ 
would allow a determination of the $\alpha_{s}$ coupling constant. 
We note also that our three predictions for the $\eta_{b}(nS)$ two-photon width 
(as well as the one from ref.\cite{Fabiano:2002nc} for 
$\Gamma_{\gamma \gamma}(\eta_b(1S))$ which is also 
based on spin symmetry) differ
significantly from a number of other predictions (see \ct{tab-res}). It
would  be also very 
interesting to have experimental indications whether or not heavy-quark spin 
symmetry could be 
broken for excited states, in view of the aforementioned possible anomaly 
for $\Gamma_{\gamma \gamma}(\eta'_c)$.


\begin{thebibliography}{99}

\bibitem{Lansberg:2006dw}
  J.~P.~Lansberg and T.~N.~Pham,
  Phys.\ Rev.\ D {\bf 74} (2006) 034001
  [arXiv:hep-ph/0603113].


\bibitem{neubert}
 M.~Neubert,
Phys.\ Rept.\  {\bf 245}, 259 (1994);
[arXiv:hep-ph/9306320].

\bibitem{nardulli}
  R.~Casalbuoni, A.~Deandrea, N.~Di Bartolomeo, R.~Gatto, F.~Feruglio and G.~Nardulli,
  Phys.\ Rept.\  {\bf 281}, 145 (1997)
  [arXiv:hep-ph/9605342].



\bibitem{DeFazio}
  F.~De Fazio, in {\it At the Frontier of Particle Physics/Handbook
of QCD}, edited by M. A. Shifman (World Scientific, 2001) 1671
[arXiv:hep-ph/0010007]~;
  arXiv:hep-ph/0010007.




\bibitem{Heister:2002if}
  A.~Heister {\it et al.}  [ALEPH Collaboration],
  Phys.\ Lett.\ B {\bf 530} (2002) 56
  [arXiv:hep-ex/0202011].

\bibitem{PDG}
  W.-M. Yao {\it et al.}  [Particle Data Group],
  J.\ Phys.\ G {\bf 33} (2006) 1.

\bibitem{Maltoni:2004hv}
  F.~Maltoni and A.~D.~Polosa,
  Phys.\ Rev.\ D {\bf 70} (2004) 054014
  [arXiv:hep-ph/0405082].


\bibitem{Swanson:2006st}
  E.~S.~Swanson,
  Phys.\ Rept.\  {\bf 429} (2006) 243
  [arXiv:hep-ph/0601110].

\bibitem{Rosner:2006jz}
  J.~L.~Rosner,
  arXiv:hep-ph/0609195.

\bibitem{Colangelo:2006aa}
  P.~Colangelo, F.~De Fazio, R.~Ferrandes and S.~Nicotri,
  arXiv:hep-ph/0609240.

\bibitem{Reinders:1982zg}
  L.~J.~Reinders, H.~R.~Rubinstein and S.~Yazaki,
  Phys.\ Lett.\ B {\bf 113} (1982) 411.


\bibitem{Dudek:2006ej}
  J.~J.~Dudek, R.~G.~Edwards and D.~G.~Richards,
  Phys.\ Rev.\ D {\bf 73} (2006) 074507
  [arXiv:hep-ph/0601137].


\bibitem{Kuhn:1979bb}
  J.~H.~Kuhn, J.~Kaplan and E.~G.~O.~Safiani,
  Nucl.\ Phys.\ B {\bf 157} (1979) 125.


\bibitem{Pham}
  T.~N.~Pham and G.~h.~Zhu,
  Phys.\ Lett.\ B {\bf 619} (2005) 313
  [arXiv:hep-ph/0412428].


\bibitem{Novikov:1977dq}
  V.~A.~Novikov, L.~B.~Okun, M.~A.~Shifman, A.~I.~Vainshtein, 
  M.~B.~Voloshin and V.~I.~Zakharov,
  Phys.\ Rept.\  {\bf 41} (1978) 1.

\bibitem{Kwong:1987mj}
  W.~Kwong, J.~L.~Rosner and C.~Quigg,
  Ann.\ Rev.\ Nucl.\ Part.\ Sci.\  {\bf 37} (1987) 325.

\bibitem{Lakhina:2006vg}
  O.~Lakhina and E.~S.~Swanson,
  Phys.\ Rev.\ D {\bf 74} (2006) 014012
  [arXiv:hep-ph/0603164].


\bibitem{Godfrey:1985xj}
  S.~Godfrey and N.~Isgur,
  Phys.\ Rev.\ D {\bf 32} (1985) 189.


\bibitem{Penin:2004ay}
  A.~A.~Penin, A.~Pineda, V.~A.~Smirnov and M.~Steinhauser,
  Nucl.\ Phys.\ B {\bf 699} (2004) 183
  [arXiv:hep-ph/0406175].



\bibitem{Brambilla:2004jw}
  N.~Brambilla, A.~Pineda, J.~Soto and A.~Vairo,
  Rev.\ Mod.\ Phys.\  {\bf 77} (2005) 1423
  [arXiv:hep-ph/0410047].



\bibitem{Mackenzie}
P. ~B. ~Mackenzie and G.~P.~Lepage,
 Phys.\ Rev.\ Lett {\bf 47} (1981) 1244 


\bibitem{Schuler:1997yw}
  G.~A.~Schuler, F.~A.~Berends and R.~van Gulik,
  Nucl.\ Phys.\ B {\bf 523} (1998) 423
  [arXiv:hep-ph/9710462].


\bibitem{Ackleh:1991dy}
  E.~S.~Ackleh and T.~Barnes,
  Phys.\ Rev.\ D {\bf 45} (1992) 232.

\bibitem{Kim:2004rz}
  C.~S.~Kim, T.~Lee and G.~L.~Wang,
  Phys.\ Lett.\ B {\bf 606} (2005) 323
  [arXiv:hep-ph/0411075].

\bibitem{Ahmady:1994qf}
  M.~R.~Ahmady and R.~R.~Mendel,
  Phys.\ Rev.\ D {\bf 51} (1995) 141
  [arXiv:hep-ph/9401315].


\bibitem{Munz:1996hb}
  C.~R.~Munz,
  Nucl.\ Phys.\ A {\bf 609}, 364 (1996)
  [arXiv:hep-ph/9601206].

\bibitem{Ebert:2003mu}
  D.~Ebert, R.~N.~Faustov and V.~O.~Galkin,
  Mod.\ Phys.\ Lett.\ A {\bf 18}, 601 (2003)
  [arXiv:hep-ph/0302044].



\bibitem{Fabiano:2002nc}
  N.~Fabiano,
  Eur.\ Phys.\ J.\ C {\bf 26} (2003) 441
  [arXiv:hep-ph/0209283].


\bibitem{yr}
N.~Brambilla {\it et al.}, {\it Heavy quarkonium physics}, CERN Yellow Report, CERN-2005-005, 
2005  Geneva : CERN, 487 pp 
[arXiv:hep-ph/0412158].

\bibitem{Lansberg:2006dh}
  J.~P.~Lansberg, Int.\ J.\ Mod.\ Phys.\ A {\bf 21} (2006) 3857
  [arXiv:hep-ph/0602091].


\bibitem{Asner:2003wv}
  D.~M.~Asner {\it et al.}  [CLEO Collaboration],
  Phys.\ Rev.\ Lett.\  {\bf 92} (2004) 142001
  [arXiv:hep-ex/0312058].

\end{thebibliography}
\end{document}